\documentclass[submission,copyright]{eptcs}
\providecommand{\event}{Fourth Workshop on Formal Methods for Autonomous Systems \\ FMAS 2022} 
\def\stateflow{Stateflow}

\def\sesf{{\scshape SeSf}}
\usepackage{cite}
\usepackage{graphicx}
\usepackage{hyperref}
\usepackage{mathtools}
\usepackage{listings}
\usepackage{multirow,tabularx}
\usepackage{verbatim}
\usepackage{caption}
\usepackage{subcaption}
\usepackage{float}
\usepackage{amssymb,amsmath,amsthm,enumitem}
\usepackage{styles/z3/style}
\usepackage{styles/proof}
\usepackage{xcolor}
\usepackage{breqn}
\usepackage{textcomp}
\usepackage{pgfplots}
\pgfplotsset{compat=1.14}

\newcolumntype{C}{>{\raggedright\arraybackslash}X}

\begin{document}
\title{Bounded Invariant Checking for Stateflow\thanks{Work partially funded by the FFI Programme of the Swedish Governmental Agency for Innovation Systems (VINNOVA) as the AVerT2 project 2021-02519.}}

\newcolumntype{s}{>{\hsize=.25\textwidth}X}
\newcommand{\definedas}{\stackrel{def}{=}}
\newcommand{\bexpr}[1]{\mathcal{SB}[\![#1]\!]}
\newcommand{\aexpr}[1]{\mathcal{SA}[\![#1]\!]}
\newcommand{\expr}[1]{\mathcal{A}[\![#1]\!]}
\newcommand\defas{\stackrel{\mathclap{\normalfont\mbox{def}}}{=}}
\newcommand{\ruleskip}{\vspace{1mm}}
\newcommand*{\bfrac}[2]{\genfrac{\[}{\]}{0pt}{}{#1}{#2}}
\setlength{\abovedisplayskip}{3pt}
\setlength{\belowdisplayskip}{3pt}
\setlength{\abovedisplayshortskip}{3pt}
\setlength{\belowdisplayshortskip}{3pt}
\renewcommand{\figurename}{Figure}

%
%
%
%
\author{Predrag Filipovikj
\institute{Scania CV AB \\ S\"o{}dert\"alje, Sweden}
\institute{KTH Royal Institute of Technology \\ Stockholm, Sweden}
\email{predrag.filipovikj@scania.com}
\and
Gustav Ung
\institute{Scania CV AB \\ S\"o{}dert\"alje, Sweden}
\email{gustav.ung@scania.com}
\and
Dilian Gurov
\institute{KTH Royal Institute of Technology \\ Stockholm, Sweden}
\email{dilian@kth.se}
\and
Mattias Nyberg
\institute{Scania CV AB \\ S\"o{}dert\"alje, Sweden}
\institute{KTH Royal Institute of Technology \\ Stockholm, Sweden}
\email{mattias.nyberg@scania.com}
}
\def\titlerunning{Bounded Invariant Checking for Stateflow}
\def\authorrunning{P. Filipovikj et al.}
\maketitle              
\begin{abstract}
\stateflow{} models are complex software models, often used as part of industrial safety-critical software solutions designed with Matlab Simulink. Being part of safety-critical solutions, these models require the application of rigorous verification techniques for assuring their correctness. In this paper, we propose a refutation-based formal verification approach for analyzing \stateflow{} models against invariant properties, based on bounded model checking (BMC). The crux of our technique is: i)~a representation of the state space of \stateflow{} models as a symbolic transition system (STS) over the symbolic configurations of the model, and ii)~application of incremental BMC, to generate verification results after each unrolling of the next-state relation of the transition system. To this end, we develop a symbolic structural operational semantics (SSOS) for \stateflow{}, starting from an existing structural operational semantics (SOS), and show the preservation of invariant properties between the two. We define bounded invariant checking for STS over symbolic configurations as a satisfiability problem. We develop an automated procedure for generating the initial and next-state predicates of the STS, and a prototype implementation of the technique in the form of a tool utilising standard, off-the-shelf satisfiability solvers. Finally, we present preliminary performance results by applying our tool on an illustrative example and two industrial models.

\end{abstract}
\section{Introduction}
\label{sec:introduction}

\stateflow{}~\cite{stateflowug} is a proprietary graphical modelling language developed and maintained by Mathworks. It is an extension of a formalism for modelling complex systems through hierarchical state machines called Statecharts~\cite{harel1987statecharts}. 
The rich graphical formalism and the variety of supporting tools in the Matlab Simulink environment enable the development of highly complex software models, which in many instances are classified as \emph{safety-critical}. The correctness of safety critical systems is regulated by domain-specific safety standards (e.g., ISO26262~\cite{iso26262} in the automotive domain), which require correct operation of such systems at all times with strongly regulated error margins. 

One way of enabling a high level of quality-assurance for safety-critical systems is to employ rigorous mathematics-based verification methods popularly known as \emph{formal verification techniques}. The main challenges of applying formal techniques for verification of \stateflow{} models stem from two main factors: i) tractability of the verification process due to the high-complexity of the \stateflow{} models, and ii) the lack of formal semantics for the \stateflow{} language publicly disclosed by Mathworks. 
The problem of formal verification of \stateflow{} models has been addressed in a number of research endeavours, which have focused either on defining a \textit{de-facto} formal semantics for the language~\cite{hamon2004operationalsemanticsstateflow,hamon2007operationalsemanticsstateflow,hamon2005denotational,bourbouh2017automatedanalysisofstateflowmodels}, or proposing a model-to-model transformation schemes for converting \stateflow{} models into some formalism of interest~\cite{yang16stateflowtransformationase,jiang2019dependablecpsstateflow,banphawatthanarak1999symbolic}. The former group of approaches often resort to exhaustive verification techniques which are likely not to scale for industrial-size models. The main limitation of the latter group of approaches is that their analysis models are not provably correct against the original \stateflow{} model. At present, industry relies mainly on the proprietary SLDV tool~\cite{hamon2008simulink} by Mathworks for the formal verification of their models. Although the tool provides a completely automated workflow for refutation-based and induction-based verification~\cite{etienne10usingsldv}, as it is proprietary, it is neither open-source nor transparent about its exact formal underpinnings and internal workings. On top of the information scarcity, the SLDV tool is distributed under a license that explicitly forbids benchmarking or any other form of direct comparison with another approach or tool, be it commercial or of purely academic nature. 

In this work, we are tackling the aforementioned challenges for formal analysis of \stateflow{} models by presenting a technique that applies \emph{bounded model checking} (BMC)~\cite{biere2003boundedmodelchecking} over \emph{symbolic executions}~\cite{king1976symbolicexecutionandtesting} of \stateflow{} models. We adopt BMC as the underlying technique for verification for two main reasons: first, to leverage the power of SAT/SMT-based model checking~\cite{barrett18smtbookchapter}, and second, to alleviate the state-space explosion by incrementally exploring all system executions of bounded length~\cite{barrett18smtbookchapter}, until the problem becomes intractable or a property violation is detected. In this paper, we focus on checking \emph{invariant} properties, which are state properties that hold in all reachable states of a given program. Even though invariant properties represent just one class of properties, based on our previous and current experiences in collaboration with industrial partners, it is often considered to be the most important one for safety-critical systems.

\paragraph{Contributions}
Our verification technique consists of the following ingredients. First, we derive a set of symbolic structural operational semantics rules (SSOS). The SSOS rules are obtained by uniformly translating into symbolic counterparts the rules of an already existing third-party SOS for \stateflow{}~\cite{hamon2004operationalsemanticsstateflow}. We build on top of this particular set of SOS rules, because it is the only available operational semantics for \stateflow{} that is suitable for our needs, and because the correctness of the rules has already been validated against the \emph{simulation semantics} of \stateflow{} (see~\cite{hamon2004operationalsemanticsstateflow}). The SSOS is needed for deriving a \textit{symbolic transition system}~(STS) at a suitably high level of granularity of the execution steps (which we choose to be the level of \stateflow{} program statements), abstracting from the intricate many-layered transitions of the original SOS. 
As our second contribution, we present two theorems that show that the SOS and SSOS \emph{simulate} each other. This result is crucial for the correctness of our technique. 
Our third contribution is a translation, using the SSOS, of \stateflow{} programs into STS over symbolic configurations, and the encoding of this STS and the given invariant property into a set of constraints in the SMT-LIB format~\cite{barret15smtlib}. This set of constraints can then be used as input to most of the modern SMT solvers. In our work, we use the Z3 SMT solver~\cite{demoura08z3} from Microsoft Research.
Finally, as our fourth and final contribution we give preliminary evidence for the practical usefulness of our approach by applying it on an illustrative \stateflow{} model. Even though initially we planned to compare our approach against the SLDV tool, in the end it was not possible due to the strict licensing constraints imposed by Mathworks. 

\paragraph{Related work} A significant portion of existing approaches for verification of \stateflow{} rely on different transformation rules and schemes for the basic \stateflow{} modeling constructs into some existing formal framework, as presented in~\cite{yang16stateflowtransformationase,jiang2019dependablecpsstateflow,chen10formalanalysisofstateflowdiagrams,meenakshi2006toolfortranslatingsimulink}. The main limitation of these approaches is the lack of means for proving the correctness of their transformation schemes, which in turn hinders the provability of the correctness of their formal models. 

Another class of approaches includes the ones that build on top of the existing \stateflow{} semantics. Miyazawa et al.~\cite{miyazawa12refinementorientedmodelsofstateflow} provide a formalization of \stateflow{} in a refinement language called Circus. The authors provide semantics characteristic to the specific refinement language, whereas in our case the semantics are defined in generalized SOS-style. The CoCoSim~framework~\cite{bourbouh2017automatedanalysisofstateflowmodels,bourbouh2020cocosim} is perhaps one of the most comprehensive bodies of work on the topic of formal verification of Simulink/Stateflow models. The framework builds on top of a denotational semantics for the Stateflow language~\cite{hamon2005denotational}. For analysis, the framework compiles the Stateflow models into Lustre models, which is the core difference to our work as we start from an SOS style semantics of the Stateflow language. Finally, there are number of approaches that treat \stateflow{} models as either hybrid or stochastic models, and apply corresponding modelling and analysis techniques and tools for verification~\cite{alur2008symbolicanalysisforcoveragesimulinkstateflow, zuliani10bayesiansimulinkstateflow,duggirala15c2e2,kaalen22stateflowstochastic}. The core difference to our approach is that these approaches treat the \stateflow{} model as either linear hybrid or Bayesian models, and resort to simulation-based techniques for the formal analysis of the model.

\paragraph{Structure}
Our paper is organised as follows. In Section~\ref{sec:background}, we outline the required background concepts that we use throughout the paper. Next, in Section~\ref{sec:symbolicsos}, we present the SSOS for \stateflow{} programs, followed by the characterization of the relationship between the concrete and symbolic semantics in Section~\ref{sec:characterization-sos-ssos}. Then, in Section~\ref{sec:transformation-and-smt-encoding}, we show how an STS over symbolic configurations can be constructed using the SSOS rules (Section~\ref{sec:stateflow-to-sts}), followed by an informal encoding procedure into an SMT-LIB script (Section~\ref{sec:sts-to-smt}). Next, we show a preliminary evaluation of our approach, based on the running example (Section~\ref{sec:evaluation-and-comparison}). 
Finally, in Section~\ref{sec:conclusions}, we present our conclusions and outline directions for future work. 

\section{Background}
\label{sec:background}

In this section, we present an overview of the concepts on which we build our work. First, in Section~\ref{sec:stateflow} we give a succinct overview of the \stateflow{} modeling language. Next, in Section~\ref{sec:h-r-sos} we give a brief overview of the existing \stateflow{} imperative language and its SOS. In Section~\ref{sec:smt-z3} we recall the general concept of Satisfiability Modulo Theories (SMT) and the Z3 tool, and finally, in Section~\ref{sec:bmc} we give an overview of Bounded Model Checking (BMC).

\subsection{\stateflow{}}
\label{sec:stateflow}

\begin{figure}[t!]
\centering 
    \includegraphics[width=0.8\columnwidth]{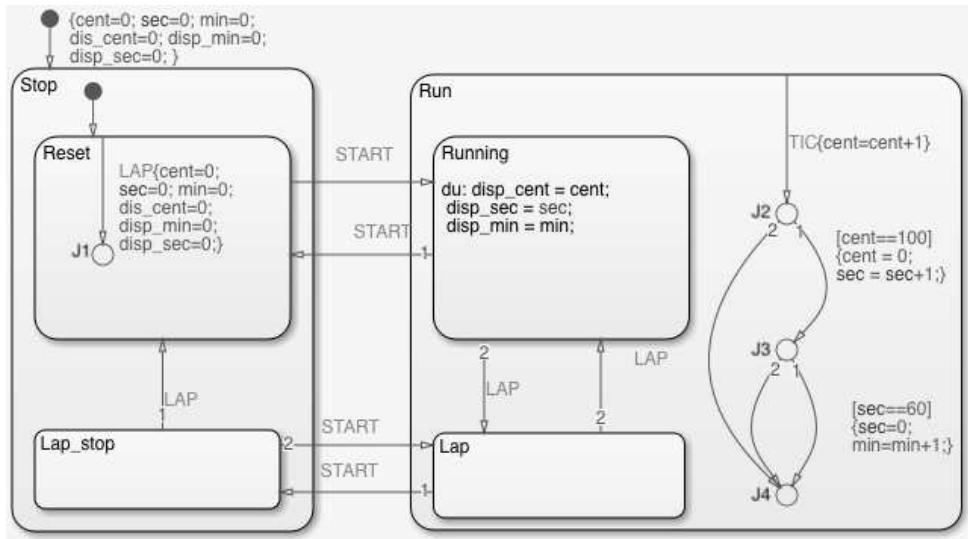} \caption{Simple \stateflow{} diagram - timer example~\cite{hamon2004operationalsemanticsstateflow}.}
    \label{fig:stateflow-scenario}
\end{figure}

\stateflow{}~\cite{stateflowug} is a graphical modeling language developed by Mathworks, integrated into the Matlab Simulink~\cite{simulinkug} modelling environment. 

A Simulink \stateflow{} model can be broadly divided into two parts: \emph{control} and \emph{data}. The control part is modeled through the concepts of \emph{Stateflow state}, \emph{connective junction}, and \emph{transition}, whereas the data part is modelled through a set of \emph{data variables} and \emph{events}. 
The control of the \stateflow{} diagram in Figure~\ref{fig:stateflow-scenario} consists of 6 Stateflow states, 4 connective junctions and 13 transitions. 
Each \stateflow{} state is decorated with a set of \emph{state actions}, which includes: 
\emph{entry ($en$)}, \emph{duration ($du$)} and \emph{exit ($ex$)}. Each action represents an atomic routine. A Stateflow state is either \emph{atomic} or \emph{composite}. Composite Stateflow states contain other states (called \emph{substates}) in their internal structure. A composite state is an Or-composition if only one of its substates can be active at any point in time, or an And-composition if there can be more than one simultaneously active states.  
The parallelism in the context of And-compositions only means concurrent activation of its substates; the execution, however, is strictly sequential and assigned by the developer. The junctions are used for modelling different branches of execution when a \stateflow{} diagram moves from one control point into another. 

The dynamics of the control flow of a \stateflow{} diagram is modelled through a set of \textit{transitions} of the following format: $\mathit{s \xrightarrow{e, c, ca, ta} s'}$, where $s$ and $s'$ are the \emph{source} and the \emph{destination} state or junction, respectively, $e$ is the \emph{transition event} that triggers the execution of the transition, which is enabled by the condition ($c$); $ca$ and $ta$ are transition actions which are executed when $c$ evaluates to true and destination is reached, respectively. 

The informal execution semantics of \stateflow{} models is very intricate and has been explained in detail in the \stateflow{} user guide published by Mathworks \cite{stateflowug}. Due to space limitations, we omit here the details of the informal execution semantics, but give in the following section an overview of a \textit{de-facto} formal one.


\subsection{\stateflow{} Imperative Language: Formal Syntax and Structural Operational Semantics}
\label{sec:h-r-sos}

In order to formalize \stateflow{}, Hamon and Rushby propose in~\cite{hamon2004operationalsemanticsstateflow,hamon2007operationalsemanticsstateflow} an imperative language that is a strict subset of the \stateflow{} graphical language. In the following, we give a brief overview of the language and its operational semantics.

The imperative language is based on the following syntactic categories: state ($s$), junction ($j$), event ($e$), action ($a$) and condition ($c$). A transition $t = (e_t, c, a_c, a_t, d)$ is composed of a transition event $e_t$, condition $c$, condition and transition actions $a_c$, $a_t$, respectively, and a destination~$d$ to which it fires. Transitions are grouped into transition lists, which ensure their sequential execution based on a predefined order. A junction definition list~$J$ associates a list of transitions with junctions. A state definition list $SD$ associates each state variable ($s$) with a state definition $sd = ((a,a,a), C, T_i, T_o, J)$. Each $sd$ contains 3 actions, a composition~$C$, lists of internal and outgoing transitions $T_i$ and $T_o$, respectively, and a junction definition list $J$. Finally, the composition~$C$ can be of type $Or(s_a, p, T, SD)$, where $s_a$ is the active state, $p$ is the path, $T$ is a transition list, and $SD$ is a list of state definitions; or of type $And(b, SD)$, which has a Boolean value~$b$ signifying whether the component is active or not, a path~$p$, and a state definition list~$SD$. In the reminder of the manuscript, we will use the term \textit{\stateflow{} program} regardless if it is modeled using the original \stateflow{} graphical language or the imperative language as we are going to be handling only models that can be rewritten in the imperative language.

The execution of a \stateflow{} program consists of processing an input event through a sequence of discrete steps. The operational semantics is formalised by a set of 27 layered rules, which precisely prescribe the sequence of actions involved in the processing of an event through the elements of the imperative language~\cite{hamon2004operationalsemanticsstateflow}. 
In our work, we refer to executions derivable using the SOS rules as \emph{concrete}. 
The based form of event processing in \stateflow{} programs expressed in the imperative language is given as: 
\begin{equation*}
    \mathit{e \vdash (P, D) \rightarrow (P', D'), tv}
\end{equation*}
~

\noindent 
which reads as follows: processing an event $e$ in an environment $D$ through a program component $P$ produces a new environment $D'$, a new program component $P'$, and a transition value $\mathit{tv}$. An environment $\mathit{D: Var \rightarrow Val}$ is a mapping from variables to values. $Env$ denotes the set of all possible environments; $P$ is an element of the \stateflow{} imperative language, whereas $\mathit{tv \in \{Fire(d,a) \: \vert \: No \: \vert \: End\}}$ is a transition value which indicates whether a transition has fired ($\mathit{Fire(d,a)}$) or not ($\mathit{No \: \vert \: End}$). All of the rules in the SOS extend and slightly differ from this general form~\cite{hamon2007operationalsemanticsstateflow}.

\begin{figure*}[t!]
\begin{subfigure}[c]{0.3\textwidth}
    \centering
        \scalebox{0.87}{
    \begin{minipage}{.3\textwidth}
    \infer{e \vdash ((e_0, c, ca, ta, d), D_1) \rightarrow{} D_2, \mathit{Fire}(d, ta)}{\begin{array}{lc}
    &(e = e_0) \lor (e_0 = \emptyset{}) \quad
    e \vdash (c, D_1) \rightarrow{} \top \\ & e \vdash (ca, D_1) \hookrightarrow D_2\end{array}}
    \end{minipage}
    }
    \caption{[t-FIRE]\textsubscript{SOS} rule}
    \label{fig:t-fire-sos-rule}
\end{subfigure}%
\hspace*{-0.7cm}
\begin{subfigure}[c]{\textwidth}
    \centering
    \scalebox{0.87}{
\begin{minipage}{\textwidth}
    \infer{\begin{array}{ll}
         & e, D_0, J, tv \vdash And\{s_0:sd_0 \cdots s_n:sd_n\} \\
         & \rightarrow{} And\{s_0:sd'_0 \cdots s_n:sd'_n\}, D_{n+1}, No
    \end{array}}{\begin{array}{cc}
         & (tv = No) \lor (tv = End)\\
         &\forall i \in [0,\dots, n] \ e, D_i, J \vdash sd_i \rightarrow{} sd'_i, D_{i+1}, No
    \end{array}}
    \end{minipage}
    }
    \caption{[AND]\textsubscript{SOS} rule}
    \label{fig:T-fire-sos-rule}
\end{subfigure}%
\caption{Illustrative sample of SOS rules.}
    \label{fig:sos-illustrative-rules}
\end{figure*}

Figure~\ref{fig:t-fire-sos-rule} illustrates the [t-Fire]\textsubscript{SOS} SOS rule. The rule describes how a \stateflow{} transition fires, and intuitively captures the following: in the concrete execution, if the evaluation of a condition evaluates to true ($\top$), and the execution of the condition action $ca$ modifies the environment, then a \stateflow{} program performs a transition, and raises a \emph{Fire} transition value. In a similar way, the Figure~\ref{fig:T-fire-sos-rule} showcases the [AND]\textsubscript{SOS} SOS rule, which describes how an And-composition is executed by sequentially executing its substates.
For the complete set of SOS rules, we refer the interested reader to the original work by Hamon~and~Rushby~\cite{hamon2004operationalsemanticsstateflow,hamon2007operationalsemanticsstateflow}.

\subsection{Satisfiability Modulo Theories and Z3}
\label{sec:smt-z3}

The problem of determining whether a Boolean formula can be made true by assigning truth values to the constituent Boolean variables is known as the \emph{Boolean satisfiability problem} (SAT). A decision procedure for SAT is a procedure that generates a (satisfying) assignment for the variables for which a given formula is true, whenever the formula is satisfiable. \emph{Satisfiability Modulo Theories} (SMT) represents an extension of SAT, where some of the logic symbols are interpreted by a background theory~\cite{barrett18smtbookchapter}. Examples of such background modulo-theories are the theory of equality, the theory of integer numbers, and the theory of real numbers. 

Z3~\cite{demoura08z3} is a state-of-the-art SMT solver and theorem prover developed by Microsoft Research. The input is a model specified in a text-based assertion language that follows the SMT-LIB standard~\cite{barret15smtlib}. Z3 provides a number of APIs for different programming languages, including C and Python, which enables the integration of the Z3 solver with other applications. The input model consists of a set of variables of specific types (also called sorts), and a set of assertions that express constraints over the variables. If the set of constraints (assertions) is satsifiable, the Z3 solver returns result \texttt{sat}, accompanied with the interpretation of the variables. In the opposite case, Z3 returns the result \texttt{unsat} and a minimal set of unsatisfiable assertions. Finally, if the model is intractable, the solver returns \texttt{unknown}.  

\subsection{Bounded Model Checking}
\label{sec:bmc}

Bounded Model Checking (BMC) is a refutation-based verification technique for checking properties over finite-state transition systems. For checking invariant properties of the type \emph{``something bad never happens"}, BMC unrolls the transition relation until one of the following becomes true: i) a ``bad" state has been reached, or ii) a predefined number of unrolling steps has been reached. The number ($k$) of unrolling steps is called \emph{bound}, while the set of all executions of length~$k$ is called \emph{reachability diameter}. Having a reachability diameter of limited size, BMC alleviates the state-space explosion problem at the expense of completeness of the procedure. 

\paragraph{Definition 1 (Symbolic Transition System)}
\label{def:sts-with-predicates}
\textit{A \emph{symbolic transition system} is a pair $S = (I, R)$, where the unary predicate~$I(\cdot)$ is a first-order logic (FOL) formula over the components of configurations representing the initial set of configurations, and the binary predicate~$R(\cdot, \cdot)$ is a formula representing the ``next-state" transition relation, satisfying the equivalences:}

\begin{align*}
    &\mathit{I(c) \: \Leftrightarrow c \in C_0}\\
    &\mathit{R(c,c')\: \Leftrightarrow (c,c') \in \ \rightarrow}
\end{align*}
~

Every initialized path in $\mathit{S}$ of length~$k$ can be characterized by the formula:

\begin{equation}
\label{eq:path-encoding}
    \mathit{path} (c_0, c_1, \ldots, c_k) \: \triangleq \: I(c_0) \land \bigwedge\limits_{i=0}^{k - 1} R(c_i,c_{i+1}), 
\end{equation}
~

\noindent and then, the existence of an initialized path of length~$k$ is equivalent to the satisfiability of the formula $\mathit{path} (x_0, x_1, \ldots, x_k)$, where $x_i$ are variables over configurations.

Let $\varphi$ be a unary predicate over configurations, i.e., a  property. We define the corresponding $k$-bounded invariant property, denoted~$\varphi^k$, as the formula:

\begin{equation}
\label{eq:k-bounded-invariance-definition}
   \forall c_0, c_1, \dots, c_k\ .\ (\mathit{path} (c_0, c_1, \dots, c_k) \Rightarrow \bigwedge\limits_{i=0}^{k} \varphi (c_i))
\end{equation}
~

\noindent A path that contains a configuration in which $\varphi^k$~does not hold is called a \emph{counter-example}, and is characterized by the negation of the above formula, i.e.:

\begin{equation}
\label{eq:counter-example-definition}
    \exists c_0, c_1, \dots, c_k.\ (path(c_0, c_1, \dots c_k) \land 
    \bigvee\limits_{i=0}^{k} \neg \varphi (c_i))
\end{equation}
~
Given that the predicates $I$, $R$, and $\varphi$ can be expressed as FOL formulas, it should be obvious how the refutation of $k$-bounded invariant properties can be reduced to an SMT problem.

\section{Symbolic Structural Operational Semantics}
\label{sec:symbolicsos}

In this section, we present our SSOS semantics for the \stateflow{} imperative language, which we use as a basis for constructing an STS $\mathit{\widehat{S}}$ for a given \stateflow{} program. We start from the existing
\textit{de-facto} SOS semantics as in~\cite{hamon2004operationalsemanticsstateflow,hamon2007operationalsemanticsstateflow}, and transform each of the SOS rules uniformly into a corresponding symbolic counterpart.

In the original formalization, the sets of variables ($\mathit{Var}$) and values ($\mathit{Val}$), as well as the sets of actions ($\mathit{Act}$) and conditions ($\mathit{Cond}$) are considered to be a part of the action language which is distinct from the \stateflow{} language itself. The details for the actions and conditions are abstracted away; however, it is assumed that the semantics of the executing actions and the evaluating conditions is available via judgments of the form: 

\begin{equation*}
    \mathit{\textnormal{(i) } e \vdash (a, D) \hookrightarrow{} D' \textnormal{ and }\textnormal{(ii) } e \vdash (c, D) \rightarrow \top \: \vert \: \bot} 
\end{equation*} 
~

\noindent
which are read as follows: (i) evaluating an action ($a$) in a current environment ($D$) produces a new environment ($D'$), and (ii) evaluating a condition ($c$) in an environment ($D$) produces either true or false Boolean value.

The set of SSOS rules is created by uniformly transforming each of the SOS rules into a corresponding symbolic rule, by: i) replacing each valuation of the program variables, called \emph{environment} ($\mathit{D}$), with a symbolic representation ($\Delta$), and ii) adding a path condition ($\mathit{pc}$). Consequently, we update the action execution and condition evaluation, which evaluate over the symbolic environment and path condition, respectively. Following the basic principles of symbolic execution~\cite{king1976symbolicexecutionandtesting}, in the set of SSOS rules we treat the data component of the language in a symbolic way, whereas the control-flow remains concrete.


\begin{figure*}[t!]
\begin{subfigure}[c]{0.3\textwidth}
    \centering
    \scalebox{0.87}{
    \begin{minipage}{\columnwidth}
        \infer{e, J \vdash (t.T, \langle\Delta_1, pc_1 \rangle) \rightarrow{} \langle\Delta_2, pc_2\rangle, \mathit{Fire}(d, ta)}{
        \begin{array}{ccc}
             &  \\
             & e \vdash (t, \langle\Delta_1, pc_1 \rangle) \rightarrow{} \langle\Delta_2, pc_2\rangle, \mathit{Fire}(d, ta)
        \end{array}
        }
    \end{minipage}
    }
    \caption{[T-FIRE]\textsubscript{SSOS} rule}
    \label{fig:t-fire-ssos-rule}
\end{subfigure}%
\hspace*{-0.7cm}
\begin{subfigure}[c]{\textwidth}
    \centering
    \scalebox{0.87}{
    \begin{minipage}{\textwidth}
    \infer{\begin{array}{cc}
         & e, \langle\Delta_0, pc_0\rangle, J, tv \vdash And\{s_0:sd_0 \cdots s_n:sd_n\} \rightarrow{} \\
         & (And\{s_0:sd'_0 \cdots s_n:sd'_n\}, \langle\Delta_{n+1}, pc_{n+1}\rangle, No)
    \end{array}}{\begin{array}{cc}
         & (tv = No) \lor (tv = End) \quad \forall i \in [0,\dots, n]\\
         & e, \langle\Delta_i, pc_i\rangle, J \vdash sd_i \rightarrow{} (sd'_i,\langle\Delta_{i+1}, pc_{i+1}\rangle, No)
    \end{array}}
    \end{minipage}
    }
    \caption{[AND]\textsubscript{SSOS} rule}
    \label{fig:and-sos-rule}
\end{subfigure}%
\caption{Illustrative sample of SSOS rules.}
    \label{fig:illustrative-rules}
\end{figure*}

We define a \emph{symbolic configuration} $\mathit{sc} \in \mathit{SC}$ as a structure $\mathit{( P, \langle} \Delta \mathit{, pc\rangle)}$, where $P$~is any component from the Stateflow imperative language. We introduce a new set of \emph{symbolic variables} (symbols), denoted $\mathit{Sym}$, and a bijection $\mathit{g: Var \rightarrow Sym}$ between the program variables and the symbols. The path condition $pc$ is simply a Boolean expression over the set of symbols, whereas the \emph{symbolic environment} $\Delta \in \mathit{SEnv}$ is a mapping $\Delta \mathit{: \mathit{Var} \rightarrow \mathit{Expr_{Sym}}}$ from program variables to (arithmetic) expressions over symbols. Finally, we assume that symbolic action execution and symbolic condition evaluation are provided via semantic functions of type $\mathcal{SA}: \mathit{Act} \rightarrow (\mathit{SEnv} \rightarrow \mathit{SEnv})$ and $\mathcal{SB}: \mathit{Cond} \rightarrow (\mathit{SEnv} \rightarrow \mathit{BExpr_{Sym}})$, respectively.

We can now define the axioms for action execution and condition evaluation, for symbolic execution of \stateflow{} programs, as follows:

\begin{equation}
\hspace*{-3mm}
\begin{aligned}
&e \vdash (a, \langle\Delta_1, pc_1\rangle) \hookrightarrow{} \langle\Delta_2, pc_1\rangle \textnormal{~if } \Delta_2 = \aexpr{a}(\Delta_1)\\
&e \vdash (c, \langle\Delta_1, pc_1\rangle) \rightarrow \langle\Delta_1, pc_2\rangle \textnormal{~if } pc_2 = pc_1 \land \bexpr{c}(\Delta_1)
\end{aligned}
\end{equation}~

\noindent 
The initial symbolic configuration is $(P, \langle \Delta_0, pc_0\rangle)$, where $P$~is a component of the \stateflow{} imperative language, $\Delta_0 = g$, and $pc_0 = \top$.

The set of SOS rules can now be uniformly translated into a corresponding SSOS counter-part. Due to space constraints, in Figure~\ref{fig:illustrative-rules}, we show two instances of the SSOS rules, which are the symbolic counter-part of the SOS rules from Figure~\ref{fig:sos-illustrative-rules}. For the complete set of SSOS rules, we refer the reader to the accompanying technical report~\cite{filipovikj21bicforstateflowprogramsarxiv}. 
The [t-FIRE] rule in Figure~\ref{fig:t-fire-ssos-rule} describes how a \stateflow{} transition ($t$) fires by appending the symbolic evaluation of the condition $t.c$ to the current path condition and by symbolically executing the condition action $t.ca$ over the current symbolic environment $\Delta$. When a transition fires, a transition event $\mathit{Fire} (t.d, t.ta)$ is generated. Similarly, the [AND] rule in Figure~\ref{fig:and-sos-rule} describes the how the And-composition is processed symbolically.   

Since we are overloading the transition relation symbol ``$\mathit{\rightarrow}$'' in the SOS and SSOS rules, further in the paper we shall use ``$\mathit{\xrightarrow[]{SOS}}$'' for transitions derivable with the SOS rules, and ``$\mathit{\xrightarrow[]{SSOS}}$'' for transitions derivable with the SSOS rules.

\section{Characterization of the SSOS}
\label{sec:characterization-sos-ssos}

Our SSOS semantics is essentially an operational semantics for symbolic execution of \stateflow{} programs. It opens up the opportunity for application of a broader spectrum of verification techniques, such as: \emph{testing} (purely symbolic, or as a combination of symbolic and concrete (concolic) testing~\cite{godefroid05dartconcolictesting}) or \emph{bounded model checking}~\cite{biere2003boundedmodelchecking}. To be able to reason symbolically over \stateflow{} programs, however, one must first provide a formal characterization of the relationship between its concrete and symbolic execution. In this section, we prove two results that characterize this relationship. In Theorem~\ref{theo:theorem-1} we show that for each derivable SSOS transition there exists a corresponding derivable SOS transition. Conversely, in Theorem~\ref{theo:theorem-2} we show that for each derivable SOS transition there exists a derivable SSOS transition.
The connection is established in both cases by means of an interpretation of the symbolic values for which the Boolean expression added to the path condition holds. 

First, we introduce some additional notation.
Let $\beta: \mathit{SEnv} \times \mathit{Env} \rightarrow \mathit{Env}$ be a function that transforms a symbolic environment~$\Delta$ into a concrete one~$\beta (\Delta, D)$ with the help of an environment~$D$ that serves as an interpretation of the symbolic values; for any $v \in \emph{Var}$, let $\beta (\Delta, D) (v)$ be defined as the value of the expression~$\Delta (v)$ in the (renamed) environment $D \circ g^{-1}$. 
Similarly, let $\mathcal{B}: \mathit{BExpr}_{\mathit{Sym}} \rightarrow (\mathit{Env} \rightarrow \mathit{Bool})$ be a function that evaluates path conditions in concrete environments, so that $\mathcal{B} [\![\mathit{pc}]\!] (D)$ is the Boolean value of the path condition~$\mathit{pc}$ in $D \circ g^{-1}$.
Finally, observing that the transitions derived by the SSOS rules only (potentially) add a conjunct to the current path condition~$pc_k$ to obtain a new path condition~$pc^{k+1}$, let $pc^{k+1}_k$ denote this added conjunct (or~$\top$, if no conjunct is added). 

\paragraph{Theorem 1.}
\label{theo:theorem-1}
\textit{If} $(P_1, \langle \Delta_1, pc_1\rangle) \xrightarrow[]{SSOS}$ $(P_2, \langle \Delta_2, pc_2\rangle , tv)$,\textit{ then for all }$D_0 \in \mathit{Env}$\textit{ s.t.} $\mathcal{B}[\![pc_1^2]\!](\beta(\Delta_1, D_0)) = \top$, \textit{ we have }$(P_1, \beta(\Delta_1, D_0))  \xrightarrow[]{SOS} (P_2, \beta(\Delta_2, D_0))$. \\ \\
Our next result establishes the reverse direction.

\paragraph{Theorem 2.}
\label{theo:theorem-2}
\textit{If} $(P_1, D_1) \xrightarrow[]{SOS} (P_2, D_2)$,\textit{ then for all }$pc_1 \in \mathit{BExpr_{Sym}}$, $\Delta_1 \in \mathit{SEnv}$ and $D_0 \in \mathit{Env}$\textit{ such that } $\beta(\Delta_1, D_0) = D_1$,\textit{ there exist }$pc_2,\; pc_1^2 \in \mathit{BExpr_{Sym}}$\textit{ and }$\Delta_2 \in \mathit{SEnv}$\textit{ such that }$pc_2 = pc_1 \land pc_1^2$, $\mathcal{B}[\![pc_1^2]\!](\beta(\Delta_1, D_0)) = \top$, $\beta(\Delta_2, D_0) = D_2$\textit{ and }$(P_1, \langle\Delta_1, pc_1 \rangle) \xrightarrow[]{SSOS} (P_2,  \langle\Delta_2, pc_2 \rangle)$. \\

\noindent For the proofs of Theorems~\ref{theo:theorem-1} and~\ref{theo:theorem-2}, we refer the reader to the accompanying technical report~\cite{filipovikj21bicforstateflowprogramsarxiv}.

There are two important corollaries of the above two results, which, for reasons of space limitations, will only be stated informally here. First, both results lift naturally to \emph{executions}, i.e., to sequences of transitions. Note in particular how in Theorem~\ref{theo:theorem-2} the ``for all~$\mathit{pc}_1$ \ldots there exists~$\mathit{pc}_2$'' part allows the sequential composition of transitions. Second, when starting from a true path condition, as one does in symbolic execution, the \emph{satisfying assignments} for the path condition at the end of any symbolic path, viewed as interpreting environments, define precisely the concrete paths that follow the symbolic one. 

Furthermore, the executions in SOS and SSOS can be shown to \emph{simulate} each other with respect to processing external events. It is well-known that invariant properties are preserved by simulation, and thus, can be checked by symbolically executing the given \stateflow{} program. Even if limited, this class of properties is important in industrial contexts, as our collaboration with Scania on formally verifying safety-critical embedded code generated from Simulink models has shown.

\section{From \stateflow{} Programs to SMT Solving}
\label{sec:transformation-and-smt-encoding}

In our work, we focus on checking invariant properties over symbolic representation of \stateflow{} programs, by means of BMC. In Section~\ref{sec:symbolicsos} we developed an SSOS for \stateflow{}, and exhibited in Section~\ref{sec:characterization-sos-ssos} a simulation relation between executions derived in SOS and SSOS, which is sufficient for the preservation of invariant properties. In the following, we show how we use the SSOS to relate \stateflow{} programs to STS over symbolic configurations. We define the \emph{k-bounded invariant checking} problem for the latter representation (Section~\ref{sec:stateflow-to-sts}), and show how this problem can be encoded as an SMT problem (Section~\ref{sec:sts-to-smt}). 

\subsection{Bounded Invariant Checking for Stateflow Programs}
\label{sec:stateflow-to-sts}

In this section, we define a version of STS that encode the \emph{symbolic} behaviors of \stateflow{} programs, and then adapt the BMC problem to such transition systems.

\paragraph{Definition 2 (STS over Symbolic Configurations)}
\label{def:sts-kappa}
\textit{A symbolic transition system over the symbolic configurations of a given \stateflow{} program is an STS $\widehat{S} = (\widehat{I}, \widehat{R})$, in the sense of Definition~\ref{def:sts-with-predicates}, but over the symbolic configurations and transitions of the program as induced by the SSOS rules.} 

\noindent $\widehat{I}(\cdot)$ and $\widehat{R}(\cdot, \cdot)$ are thus a unary ``initialization'' predicate and a binary ``next-state" predicate over the symbolic configurations of the program, respectively.

The formal relationship between an STS over symbolic configurations~$\widehat{S}$ and an ordinary STS~$S$ of a \stateflow{} program is given by the following result.
\paragraph{Proposition 1.}
\label{theo:sts-sts-kappa-relationship}
\textit{Let SF be a \stateflow{} program, $S = (I, R)$ be an STS over its concrete configurations as induced by the SOS rules, and $\widehat{S} = (\widehat{I}, \widehat{R})$ be an STS over its symbolic configurations as induced by the SSOS rules. Then, the following equivalences hold:} \\

{ \small
$(1)~~\widehat{I}(P, \langle \Delta, pc \rangle) \>\Leftrightarrow\> \exists D_0 \in \mathit{Env}.\ I(P, D_0) \>\land\> I(P, \beta(\Delta, D_0)) \>\land\> \mathcal{B}[\![pc]\!](\beta(\Delta, D_0)) $ \\

$\begin{array}{lll}(2)~~\widehat{R}((P, \langle  \Delta_1, pc_1  \rangle), (P', \langle  \Delta_2, pc_2  \rangle)) \> \Leftrightarrow & \exists D_0 \in \mathit{Env}.\ \mathcal{B}[\![pc_1]\!](\beta(\Delta, D_0)) \land \mathcal{B}[\![pc_2]\!](\beta(\Delta', D_0)) \> \\
& \land \ R((P, \beta(\Delta_1, D_0)), (P', \beta(\Delta_2, D_0)))) \end{array}$
}
\begin{proof}
Follows from Definition~\ref{def:sts-with-predicates} (see Section~\ref{sec:bmc}), and from Definition~\ref{def:sts-kappa}, Theorem~\ref{theo:theorem-1} and Theorem~\ref{theo:theorem-2} (see Section~\ref{sec:characterization-sos-ssos}). The complete proof can be found in the full version of the manuscript~\cite{filipovikj21bicforstateflowprogramsarxiv}.
\end{proof}

Now, let~$\varphi$ be a predicate over the concrete configurations of a \stateflow{} program. Predicate $\varphi$ induces a corresponding predicate $\widehat{\varphi}(sc) \triangleq \varphi(sc[g^{-1}])$ over the symbolic configurations $\mathit{sc = (P, \langle} \Delta \mathit{, pc \rangle)}$, where~$g$ is the bijection from Section~\ref{sec:symbolicsos}. Assuming an interpretation for the \emph{path} and \emph{k-bounded invariant property} formulas for executions over symbolic configurations, the counter-example path formula~(\ref{eq:counter-example-definition}) for symbolic executions can be rewritten as follows:
\begin{equation}
\label{eq:counter-example-definition-sts-kappa}
    \exists sc_0, \dots, sc_k.\ (path(sc_0, \dots, sc_k) \land 
    \bigvee\limits_{i=0}^{k} \neg\widehat{\varphi}(sc_i))
\end{equation}~
\noindent
Based on formula~(\ref{eq:counter-example-definition-sts-kappa}), we derive the following. 
\paragraph{Theorem 3.}
\label{theo:sts-kappa-bmc}
\textit{Let SF be a \stateflow{} program, $\widehat{S} = (\widehat{I}, \widehat{R})$ be an STS over its symbolic configurations, and $\varphi^k$ be a k-bounded invariant property. Then, the following two statements are equivalent:}

\begin{enumerate}
 \item \textit{SF satisfies the k-bounded invariant property $\varphi^k$.}
 \item \textit{The formula $path(sc_0, \dots sc_k) \land \bigvee\limits_{i=0}^{k} \neg\widehat{\varphi}(sc_i)$ is UNSAT.}
\end{enumerate}

\begin{proof}(By contradiction.)
Assume that a given \stateflow{} program does not satisfy the \emph{k-}bounded invariant property~$\varphi^k$, and that statement (2) holds. By the definition of a \emph{k}-bounded invariant property, such a property fails if there exists a path in which the last configuration violates~$\varphi$. By Definition~\ref{eq:counter-example-definition-sts-kappa}, such a path exists if the formula given in (2) is satisfiable (SAT), which contradicts the initial assumption. The other direction is shown analogously. 
\end{proof}

Now that we have formally defined BMC invariant checking for STS over symbolic configurations, in the next section we show how to construct such STS for a given \stateflow{} program. 

\subsection{From \stateflow{} Programs to SMT Scripts}
\label{sec:sts-to-smt}

In this section, we present a succinct version of a procedure for deriving an STS from a given \stateflow{} program using the set of SSOS rules, and the transformation of the STS predicates into quantifier-free FOL formulas that can be used for \emph{k-bounded invariant checking} over symbolic configurations, as defined in Theorem~\ref{theo:sts-kappa-bmc}. 

The procedure presented in this section demonstrates the derivation of an STS in which the transitions between configurations correspond to transitions at the top \texttt{Or}-composition level, which corresponds to our Stopwatch running example (see Figure~\ref{fig:stateflow-scenario}). Due to the layered structure of the imperative language, each such transition consists of a series of transitions corresponding to the various constituent syntactic components of a given Or-composition. Our approach to the derivation of the top-level transitions is to use our SSOS to perform \emph{symbolic execution} between any possible pair of consecutive control points of the program, for arbitrary data values. One should note that in general case, the derivation of the STS is not strictly bound to the top-level component, as it can be done against any syntactic class of the \stateflow{} imperative language.

As a result of our adopted modeling principle, the configurations for the induced STS are of the following type: $(Or, \langle \Delta, pc\rangle)$. Even though the program component during execution remains the same (the top-level Or-component), it can be the case that its internal configuration changes. The internal configuration of an Or-component is characterized by the set of active substates. Consequently, the \emph{program control points} correspond to the possible internal configurations at the top Or-composition level.

We model the Stateflow program control points using a set of Boolean variables, denoted as $\mathit{Var_C}$. For every~$\mathit{Or}$ control point, $\mathit{Var_C}$ can be partitioned into two subsets: the set $\mathit{Var_{C^+} = \{v \: | \: v \in \mathit{Var_C}, \: v = \top\}}$ corresponding to the active states of~$\mathit{Or}$, and $\mathit{Var_{C^-}} = \mathit{Var_C} \setminus \mathit{Var_{C^+}}$. Thus, a control point~$\mathit{Or}$ is characterized by the formula:
\begin{equation}
\label{eq:orphi}
    \Phi_{\mathit{Or}} \>\triangleq\> \bigwedge\limits_{v \in Var_{C^+}} v \; \land \bigwedge\limits_{v \in Var_{C^-}} \neg v
\end{equation}~

The path condition~$pc$ is a quantifier-free Boolean expression over symbols, and as such can be viewed as a quantifier-free FOL formula $\Phi_{pc}$. Based on~$\Delta$, one can construct a quantifier-free FOL formula modulo theory of arithmetic for~$\Phi_\Delta$, over the set of data variables $\mathit{Var_D} = \mathit{Var} \setminus \mathit{Var_C}$ as follows: 

\begin{equation}
\label{eq:deltaphi}
\begin{aligned}
    \Phi_\Delta \triangleq \bigwedge\limits_{v \in \mathit{Var_D}} v' = \Delta(v)
\end{aligned}
\end{equation}~

Now that we have defined the construction of quantifier-free FOL formulas for each of the components of the symbolic configurations of an STS, we can construct, for every transition~$T_i$ between symbolic configurations, a quantifier-free FOL formula~($\Phi_{T_i}$) modulo theory of arithmetic, as follows:

\begin{equation}
\label{eq:transitionphi}
    \Phi_{T_i} \>\triangleq\> \Phi_{Or} \land \Phi_{pc_1^2} \Rightarrow \Phi_{Or'} \land \Phi_{\Delta'}
\end{equation}~

\noindent
Intuitively, the formula~(\ref{eq:transitionphi}) can be interpreted as follows: when the program is at control point~$Or$, the path condition~$pc_1^2$ gives the condition for the program to move to the control point~$Or'$, upon which the data will change according to~$\Delta'$. 

Finally, based on the formula~(\ref{eq:transitionphi}) and Proposition 1, we encode the predicates $\widehat{I}$~and~$\widehat{R}$ as the following quantifier-free FOL modulo theory of arithmetic formulas:

\begin{equation}
\label{eq:i-and-r-fol}
    \begin{aligned}
        &\widehat{I} \ \triangleq \ \Phi_{\mathit{Or}_\emptyset} \land \Phi_{\Delta_0} \\
        &\widehat{R} \ \triangleq \ \bigwedge\limits_{T_i \in T} \Phi_{T_i}
    \end{aligned}
\end{equation}
~

\noindent
where $T$ is the set of all derivable SSOS transitions from the initial top-level composition, which can be computed using any search algorithm starting from the initial program control point ($Or_\emptyset$). 

The final step in the process of generating an SMT model is the encoding of the FOL-formulas into corresponding SMT assertions. For details on how the FOL formulas are encoded into SMT assertions, we refer the readers to the full version of this manuscript~\cite{filipovikj21bicforstateflowprogramsarxiv}.

\section{Implementation and Experimental Comparison}
\label{sec:evaluation-and-comparison}

In this section, we first present an implementation of our approach, henceforth referred to as the \sesf{} tool. Even though the most natural way to assess the applicability and the practical usefulness of our approach is to benchmark it against the SLDV tool on a wider set of use cases, in the end it was not possible due to the licensing constraints described in Section~\ref{sec:introduction}. Therefore, we proceed with benchmarks only with our own tool. As benchmarks, we use three Stateflow models, including the Stopwatch running example from Section~\ref{sec:stateflow}, and two industrial models provided by Scania. The main purpose of the experimental comparison is to assess how \sesf{} performs w.r.t. execution time and scalability.

\paragraph{Implementation}

The \sesf{} tool is composed of two main components: i) a symbolic execution engine that generates the STS in terms of its constituent predicates $\widehat{I}$ and $\widehat{R}$, and ii) an SMT-based model-checking engine which translates the $\widehat{I}$ and $\widehat{R}$ predicates into corresponding SMT formulas, and performs their unrolling alongside the user-provided invariant property (incremental or fixed), either until a predefined bound is reached, or until the problem becomes intractable. The tool is written in Python, and part of the source code is already publicly available~\cite{sesfGitLink}.

In order to be able to analyze the model, we need to transform it into a formal counter-part which is suitable for analysis. For this purpose, we first measure the preprocessing time of the tool, i.e., the time required for deriving an analysis model from the Stopwatch Stateflow model. Our \sesf{} tool requires 5 seconds for generating the STS, translating it into an SMT script, and unrolling it for 200 execution steps, without performing syntax and consistency check. Note that \sesf{} performs the unrolling step as part of the model generation. Therefore, the model construction time is dependent on the unrolling bound. Another approach would be to perform unrolling during verification step on an as needed basis. This ensures that no unnecessary unrolling is performed.

In our first benchmark, we apply \sesf{} on the Stopwatch model with the following parametric invariant property: \emph{``The value of cent is always between 0 and X''}, for $X \in \{25, 50, 75, 98\}$. By inspecting the Stopwatch model one can see that the variable $cent$ gets values in the interval $[0, 99]$, thus for all values of~$X$ there is a counter-example. In this case \sesf{} was able to find a counter-example for each instance of the property, within the following time in seconds for each value of $X (cent)$: \sesf{} = \{7, 9, 17, 31\}.

Next, we analyze the Stopwatch model against the following invariant property: \emph{``The value of sec is always between 0 and 1''}, given a reachability diameter of [100, 125, 150, 175, 200] execution steps. For the aforementioned set of reachability diameters, the \sesf{} terminates in [47, 99, 181, 319, 487] seconds verification time, and negligible model construction time in all cases. A counterexample is found for the last reachability diameter.

The first industrial \stateflow{} model is a part of a larger vehicle feature that performs identification of a new driver. The model is composed of 6 innermost states, and 29 transitions. It differs from the Stopwatch model in that it is driven by input variables, and not by events. Therefore, the input variables were considered to be free variables in each time step. \sesf{} could analyze a true invariant property for this model with a diameter of 50 in roughly 4.5s.  

The second industrial model is used to set program variables based on engine states, and is mostly composed of junctions and transitions between them. To make the analysis easier, we developed a synthetic model with $n=5$ junctions and $2^n$ transitions. On this particular model, \sesf{} did not terminate within 2 hours. Our hypothesis for such a poor performance is that \sesf{} can only check the property against a model derived after a complete execution step at the top level instead of after each syntactic element is processed. Fortunately, this limitation of \sesf{} is of a purely implementation nature that will be fixed in future releases, and does not affect the formal underpinning of our approach. To better understand the importance of this limitation, we inspected 72 industrial models, and we discovered that only 2 of them use junction-based sub-parts.

\section{Conclusion}
\label{sec:conclusions}

We presented a technique for provably correct symbolic analysis of \stateflow{} programs with respect to invariant properties using BMC. To this end, we developed a symbolic structural operational semantics (SSOS) for the \stateflow{} language based on the previous work by Hamon and Rushby~\cite{hamon2004operationalsemanticsstateflow,hamon2007operationalsemanticsstateflow}. 
We characterized the relationship between the two semantics by exhibiting a simulation relation between them. Next, we defined the bounded invariant checking problem for STS over symbolic configurations, as induced for a given \stateflow{} program by the set of the SSOS operational rules, and presented informally a procedure for deriving the initial and next state predicates of the STS. Finally, we showed how to generate, from the STS, a set of quantifier-free FOL assertions in SMT-LIB format suitable for analysis using state-of-the-art SMT solvers. The main benefit of our work is that it lays down the foundations for the development of tools for the scalable verification of complex industrial \stateflow{} models by means of existing symbolic techniques, which we demonstrated with bounded invariant checking on several use-case models. Even though we initially planned to compare our approach against the state-of-the-practice SLDV tool, we had to withdraw from our idea once we discovered the license constraints imposed by Mathworks.

Our work can be extended in several directions. First, our formal characterization of the SSOS can be strengthened by means of a stronger equivalence between the concrete and symbolic representations of \stateflow{} programs, to formally underpin the symbolic verification of a wider class of properties than invariant properties, such as LTL properties. 
Second, one can explore the possibility of extending our BMC approach from refutation-based to a verification one, by adding induction~\cite{demoura03BMCrefutationtoverification}. Along this line of research, one could include the option of converting the generated STS into an input format for tools that implement more sophisticated model checking algorithms, such as Lustre~\cite{pilaud1987lustre} models for the Kind2~model~checker~\cite{champion16kind2}. 
Finally, one can extend our experimental evaluation in terms of the number of models, and include other tools in our comparison, such as the CoCoSim framework.

\bibliographystyle{eptcs}
\bibliography{references}
\end{document}